\documentclass[%
aps,prmaterials,twocolumn,superscriptaddress,floatfix,
reprint,
amsmath,amssymb,
longbibliography
]{revtex4-2}

\usepackage{graphicx}
\usepackage{dcolumn}
\usepackage{bm}
\usepackage[mathlines]{lineno}
\usepackage{xcolor}
\usepackage{booktabs}
\usepackage{multirow}
\usepackage{physics}
\usepackage{siunitx}
\usepackage{hyperref}

\begin{document}

\title{Alloying effects on the interface cohesion of grain boundaries in the Ni--Fe--Cr system}

\author{A. F. Baumann}
\email{aljoscha.baumann@iwm.fraunhofer.de}
\affiliation{Fraunhofer IWM, W\"ohlerstra\ss e 11, 79108 Freiburg, Germany}
\author{R. Ghassemizadeh}
\affiliation{Fraunhofer IWM, W\"ohlerstra\ss e 11, 79108 Freiburg, Germany}
\author{D. F. Urban}
\affiliation{Fraunhofer IWM, W\"ohlerstra\ss e 11, 79108 Freiburg, Germany}
\affiliation{Freiburg Materials Research Center (FMF), Albert-Ludwigs-University of Freiburg, Stefan-Meier-Stra{\ss}e 21, 79104 Freiburg, Germany}

\begin{abstract}
Grain boundaries (GBs) in polycrystalline Ni-based alloys strongly influence mechanical strength and fracture behavior, yet the effect of alloying elements such as Fe and Cr on GB cohesion remains difficult to quantify systematically. Using density functional theory (DFT), we compute rigid-body separation (RBS) curves for six symmetric-tilt Ni GBs, extracting the work of separation $W_{\mathrm{sep}}$ and maximum traction $\sigma_{\max}$ via the universal binding energy relation (UBER). We find that substitution of Cr or Fe into the GB plane increases both quantities relative to pure Ni GBs. To overcome the size limitations of DFT, we compare several machine-learning interatomic potential (MLIP) foundation models against the DFT dataset and identify a model achieving quantitative agreement with DFT for $W_{\mathrm{sep}}$ and $\sigma_{\max}$. This MLIP enables systematic sampling of alloying configurations at finer concentration steps. We show that alloying-induced changes in GB cohesion depend on both, element type and local structures, revealing trends that are inaccessible with DFT alone. Extending the analysis to a ternary $\mathrm{Ni}_{56}\mathrm{Fe}_{20}\mathrm{Cr}_{24}$ composition, inspired by the principal constituents of the widely used commercial Alloy~718, shows that the binary trends provide a useful basis for interpreting the combined effect of both substitutes, while explicit ternary calculations are required to capture their interplay quantitatively. Together, these results provide a validated workflow for quantifying the effect of alloying elements on GB cohesion in multicomponent alloys. The resulting values may serve as input for continuum-scale fracture models, paving the way toward atomistically informed multiscale modeling.
\end{abstract}

\maketitle

\section{Introduction}\label{sec:intro}

Nickel-based alloys are a cornerstone of modern high-performance materials, owing to their ability to maintain mechanical integrity and chemical stability under extreme operating conditions \cite{liuMicrostructureManipulationHightemperature2025,moverareReviewRecentAdvances2025}. They are widely deployed in energy conversion systems, chemical processing, and aerospace applications, where high temperatures, complex stress states, and aggressive environments are routinely encountered \cite{gudivadaRecentDevelopmentsNickelbased2023,salehAdditiveManufacturingNickel2026}. These multicomponent Ni-based alloys provide an ideal platform for investigating the interplay between chemistry, microstructure, and mechanical response. In particular, ternary Ni--Fe--Cr alloys are frequently used as model systems for oxidation- and corrosion-resistant materials and serve as a simplified representation of more complex commercial alloys, such as Alloy 718 \cite{kangReviewCorrosionBehaviour2023,karimihaghighiEffectAlloyingElements2023,singhAdvancesCorrosionHighTemperature2024, hosseiniReviewMechanicalProperties2019}. The Ni--Fe--Cr system is employed here as a representative, technologically relevant test case for developing and validating atomistic methodologies applicable to complex alloys and their interface mechanics.

A key microstructural feature governing the performance of Ni-based alloys is their grain boundaries (GBs). GBs strongly influence mechanical strength, fracture behavior, and environmental degradation, often acting as preferential sites for crack initiation and accelerated transport of detrimental species \cite{dehmImplicationGrainboundaryStructure2022,krauseReviewGrainBoundary2019a, janischSegregatedLightElements2003,kruppDynamicEmbrittlementTime2005}. Accordingly, a quantitative understanding of GB cohesion and separation is essential for establishing physically grounded links between atomistic interfacial mechanisms and continuum-scale models of damage evolution and macroscopic failure \cite{liuGrainBoundaryEnergy2025,uruchidaComputationalApproachGrain2024}.

At the atomistic scale, GB fracture can be studied using rigid-body separation (RBS) simulations, in which the total energy of a simulation cell containing the GB is evaluated as the two adjoining grains are incrementally and rigidly displaced normal to the interface, ultimately forming two free surfaces \cite{tahirHydrogenEmbrittlementCarbon2014,maiHighthroughputInitioStudy2025}. The resulting energy--separation data can be described using the universal binding energy relation (UBER), a general framework for modeling the energy as a function of displacement between two atomic planes \cite{roseUniversalBindingEnergy1981,banerjeaOriginsUniversalBindingenergy1988,janischInitioTensileTests2010}. In the following, all energies entering the UBER expression are normalized by the interface area $A$. For a relative displacement $d$ of the two grains normal to the interface, measured from the initial relaxed configuration,

\begin{equation}
    E(d) = E_\infty + E_b \left( 1 + \frac{d - d_0}{L} \right) \exp\left[ -\frac{d - d_0}{L} \right].
    \label{eq:uber}
\end{equation}
Here, $E_\infty$ is the energy of the fully separated system, i.e., two independent free surfaces ($d \to \infty$), $d_0$ is the equilibrium separation, $E_b$ is the binding energy, i.e., the energy at the minimum of the curve relative to the fully separated state, and $L$ is a characteristic length scale controlling the curvature of the energy--displacement curve. Throughout the remainder of this paper, we report the work of separation $W_{\mathrm{sep}} = -E_b$, i.e., the energy per unit area required to separate the two grains into two free surfaces, so that larger values indicate stronger cohesion. The corresponding traction, which quantifies the interfacial force per unit area as a function of separation, is obtained by differentiating the energy with respect to $d$:

\begin{equation}
    \sigma(d) = \dv{E}{d} = \frac{W_{\mathrm{sep}}}{L} \frac{d - d_0}{L} \exp\left[ - \frac{d - d_0}{L} \right].
    \label{eq:uber_derivative}
\end{equation}

The maximum of $\sigma(d)$ represents the maximum cohesive traction $\sigma_{\max}$ of the interface, i.e., the ideal decohesion strength under rigid separation. It occurs at $d = d_0 + L$ and is given by

\begin{equation}
    \sigma_{\max} = \frac{W_{\mathrm{sep}}}{eL}.
    \label{eq:uber_max_traction}
\end{equation}

\begin{figure}
\includegraphics[width=0.7\columnwidth]{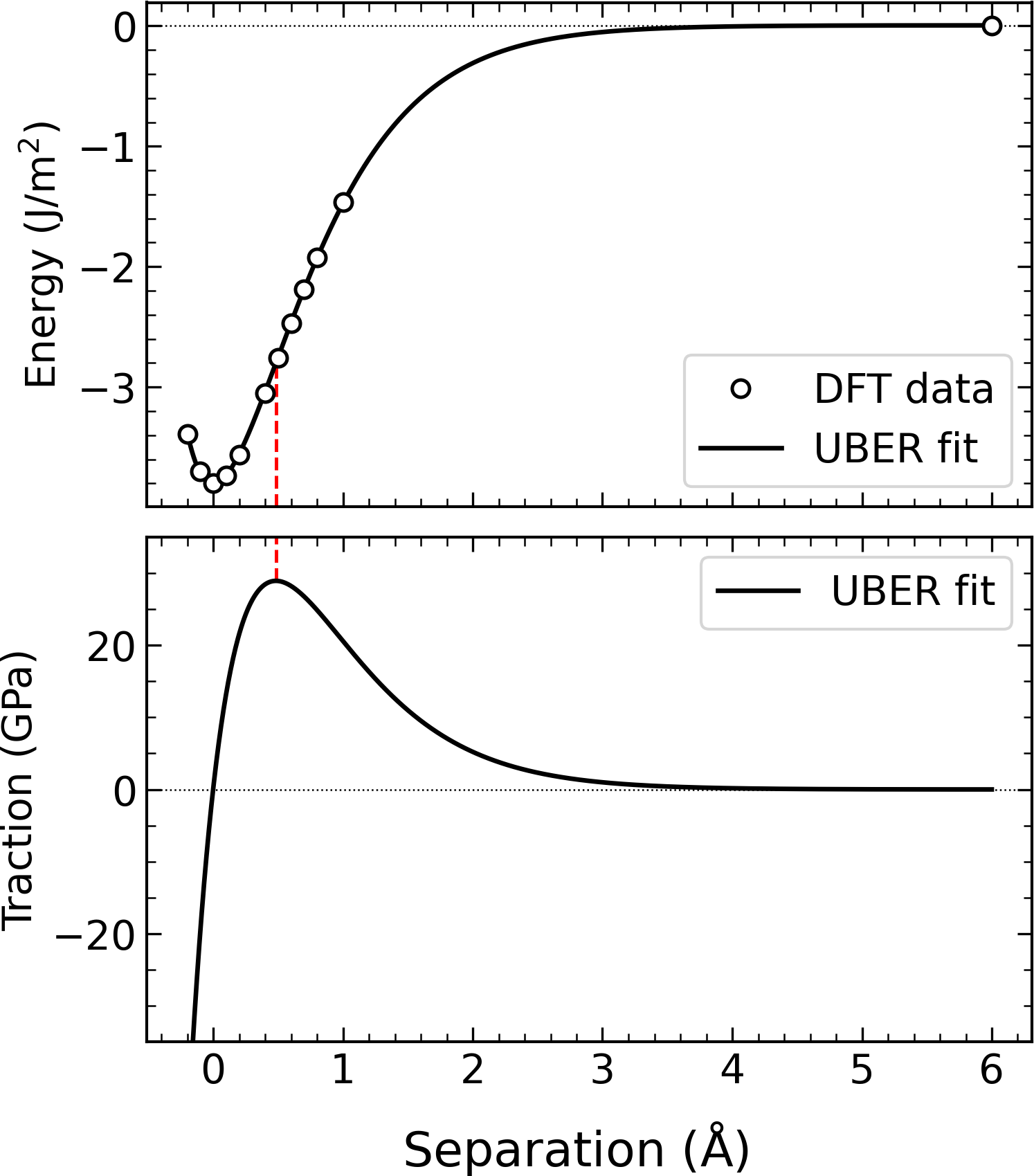}
\caption{Universal binding energy relation curve for the $\Sigma3~(111)~[\bar{1}10]$ Ni grain boundary. Top: energy as a function of separation distance, with DFT data points (open circles) and the corresponding UBER fit (solid line), referenced to the energy at the largest separation (6~\AA{}). Bottom: the associated traction, obtained as the derivative of the fitted energy curve, with the maximum traction $\sigma_{\max}$ indicated by the dashed vertical line.}
\label{fig:uber_theo}
\end{figure}

Thus, the maximum traction depends on both the work of separation and the characteristic length $L$. Figure \ref{fig:uber_theo} shows the result of a typical RBS simulation. The upper panel shows the energy--separation data, to which Eq.~\eqref{eq:uber} is fitted, with energies referenced to the energy at the largest separation. The depth of the minimum of this curve corresponds to $W_{\mathrm{sep}}$, with a separation of 6~\AA{} taken to represent two fully separated, free surfaces. The lower plot shows the derivative of the fitted curve, and the maximum traction $\sigma_{\max}$ is indicated. The quantities obtained from this procedure, such as $W_{\mathrm{sep}}$ and $\sigma_{\max}$, can be used to inform traction--separation laws for cohesive-zone models in finite-element method (FEM) simulations \cite{mollerFractureInitioForcebased2018,tahirInitioCalculationTraction2013}.

Density functional theory (DFT) provides a reliable description of interfacial cohesion at the electronic-structure level and, as a first-principles method, is not restricted to a specific chemical subspace, making it a powerful tool for describing multicomponent materials and complex local structural motifs \cite{razumovskiyFirstprinciplesaidedDesignNew2015,janischInterstitialImpuritiesGrain2009}. However, its computational cost severely restricts the accessible system size and configurational sampling. This limitation is especially problematic for multicomponent alloys, where realistic grain boundaries exhibit chemical disorder, segregation, and structural complexity that require large simulation cells far beyond what is feasible with DFT.

To address these challenges, we employ a machine-learning interatomic potential (MLIP) based on the Graph Atomic Cluster Expansion (GRACE) framework \cite{bochkarevGraphAtomicCluster2024}. Within this framework, several \emph{foundation models} are available, which have been trained on large datasets of DFT calculations, such as OMAT \cite{barros-luqueOpenMaterials20242026} or Alexandria \cite{cavignacAIDrivenExpansionApplication2026}, and which differ in complexity and the training dataset used \cite{lysogorskiyGraphAtomicCluster2026}. Using our own DFT reference data, we select the foundation model that achieves the best agreement with the DFT results. The resulting MLIP enables an accuracy comparable to DFT while allowing simulations of large, chemically complex GB structures. By combining GRACE with RBS calculations, we demonstrate how MLIPs can be used to systematically study GB cohesion in multicomponent alloys at scales inaccessible to first-principles methods. The workflow presented here, which includes calculating DFT reference data, performing large-scale GB separation simulations with a MLIP, and extracting energy--separation data, is fully transferable to a wide range of alloy systems and establishes a framework for linking atomistic modeling to continuum-scale fracture simulations.

The remainder of this paper is organized as follows: Sec.~\ref{sec:theo} describes the construction of the grain boundary structures and the computational details of the DFT and MLIP calculations. Sec.~\ref{sec:results} presents the results and discussion, beginning with the properties of the Ni grain boundaries (Sec.~\ref{subsec:gbwidening}), followed by the DFT-based chemical effects of Cr and Fe substitution (Sec.~\ref{subsec:dft_res}) and the comparison of GRACE foundation models to DFT data (Sec.~\ref{subsec:mlip_parity}). Using the selected MLIP, we then analyze the configurational effects of binary Cr and Fe substitution in large simulation cells (Sec.~\ref{subsec:config_effect}) and extend this analysis to a ternary $\mathrm{Ni}_{56}\mathrm{Fe}_{20}\mathrm{Cr}_{24}$ composition (Sec.~\ref{subsec:ternary}). Finally, Sec.~\ref{sec:conc} summarizes our findings.

\section{Theoretical approach}\label{sec:theo}

\subsection{Grain boundary structures}\label{subsec:strucs}

\begin{figure*}
\includegraphics[width=0.9\textwidth]{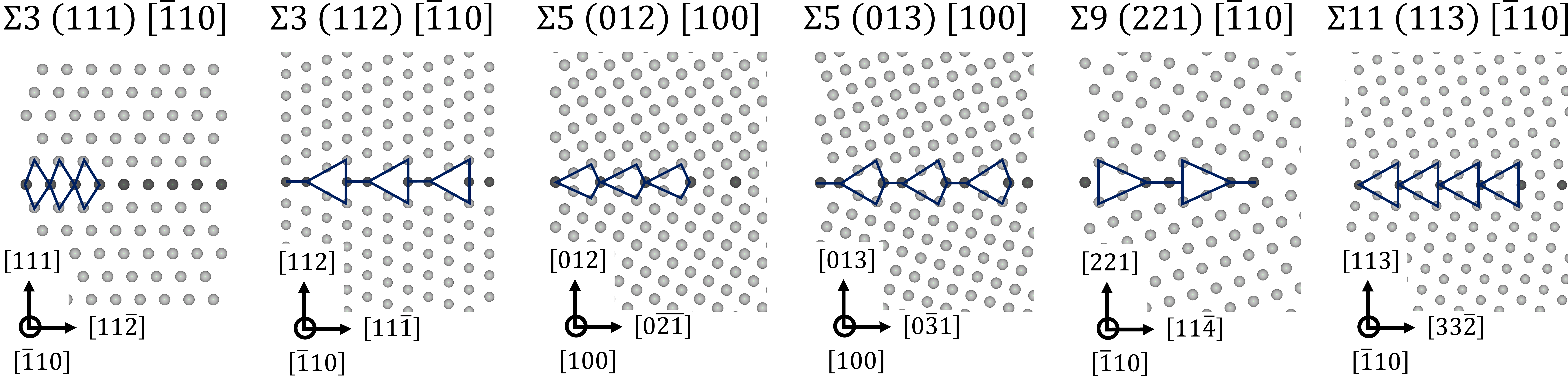}
\caption{Grain boundary structures are shown with the corresponding tilt axis oriented perpendicular to the plane of the page. For clarity, the repeating structural motifs of the grain boundaries are highlighted. Atoms depicted in darker shading are located within the grain boundary plane.}
\label{fig:gb_strucs}
\end{figure*}

First, an energy--volume curve was computed and fitted to the Birch--Murnaghan equation of state \cite{birchFiniteElasticStrain1947,murnaghanCompressibilityMediaExtreme1944} to determine the equilibrium lattice constant of ferromagnetic Ni, yielding a value of 3.518~\AA{}, in good agreement with the literature \cite{haynesCRCHandbookChemistry2016,daveyPrecisionMeasurementsLattice1925,zhuDensityFunctionalTheoryBased2025}. Using this lattice constant, symmetric-tilt grain boundary structures were constructed with the \textit{atomsk} software \cite{hirelAtomskToolManipulating2015}. For each grain boundary, face-centered cubic (fcc) Ni was oriented with the tilt axis along $x$ and the grain boundary normal along $z$. Following the approach reported by Zheng \textit{et al.}~\cite{zhengGrainBoundaryProperties2020}, pairs of Ni atoms separated by less than 1.5~\AA{} were replaced by a single Ni atom at their midpoint. The resulting simulation cells contained two grain boundaries, located at fractional coordinates $z = 0$ and $z = 0.5$. The height of the two grains was chosen such that the $k$-point density used for the entire cell (see Sec.~\ref{subsec:comp_detail}) resulted in a single $k$-point along the grain boundary normal ($c > 42$~\AA{}). This procedure was applied to all six investigated grain boundaries: $\Sigma3~(111)~[\bar{1}10]$, $\Sigma3~(112)~[\bar{1}10]$, $\Sigma5~(012)~[100]$, $\Sigma5~(013)~[100]$, $\Sigma9~(221)~[\bar{1}10]$, and $\Sigma11~(113)~[\bar{1}10]$, whose structures are shown in Fig.~\ref{fig:gb_strucs}. The dimensions of the supercells are reported in Table~\ref{tab:supercells}. 

These GBs cover a wide range of GB energies, which are defined as:

\begin{equation}
    \gamma_{\mathrm{GB}}
    = \frac{
        E_{\mathrm{GB}}
        - N_{\mathrm{GB}} E_{\mathrm{bulk}}^{(\mathrm{at})}
    }{2A},
    \label{eq:gb_energy}
\end{equation}
where $E_{\mathrm{GB}}$ is the total energy of the relaxed grain boundary supercell, $N_{\mathrm{GB}}$ is the number of atoms in that supercell, $E_{\mathrm{bulk}}^{(\mathrm{at})}$ is the reference energy per atom of the undisturbed fcc-Ni bulk crystal, $A$ is the cross-sectional area of the grain boundary plane, and the factor of~$2$ accounts for the two equivalent grain boundaries contained in the periodic slab cell.

\begin{table}[htbp]
\centering
\caption{Lattice parameters $a$, $b$, $c$, angles $\alpha$, $\beta$, $\gamma$, and number of atoms $N$ of the initial GB supercells used in the DFT calculations.}
\label{tab:supercells}
\begin{tabular}{lccccccc}
\toprule
Grain boundary & $a$ (\AA{}) & $b$ (\AA{}) & $c$ (\AA{}) & $\alpha$ ($^\circ$) & $\beta$ ($^\circ$) & $\gamma$ ($^\circ$) & $N$ \\
\midrule
$\Sigma3~(111)~[\bar{1}10]$ & 2.49 & 2.49 & 48.75 & 90 & 90 & 60 & 24 \\
$\Sigma3~(112)~[\bar{1}10]$ & 2.49 & 6.09 & 42.85 & 90 & 90 & 90 & 58 \\
$\Sigma5~(012)~[100]$ & 4.31 & 4.31 & 47.21 & 90 & 90 & 48.19 & 60 \\
$\Sigma5~(013)~[100]$ & 3.52 & 5.56 & 43.96 & 90 & 90 & 90 & 78 \\
$\Sigma9~(221)~[\bar{1}10]$ & 2.49 & 7.46 & 42.22 & 90 & 90 & 90 & 70 \\
$\Sigma11~(113)~[\bar{1}10]$ & 4.31 & 4.31 & 46.68 & 90 & 90 & 33.56 & 44 \\
\bottomrule
\end{tabular}
\end{table}

For each grain boundary plane orientation, a corresponding simulation cell was oriented such that separation of bulk Ni along the normal to this plane could be computed analogously to the GB RBS calculations. Three chemical configurations were considered for each GB in the DFT calculations: the pure Ni GB and two substituted GBs with either Cr (Cr@GB) or Fe (Fe@GB) replacing all Ni atoms in one of the two GB planes of the simulation cell. For the calculations with the GRACE MLIP, supercells were constructed by replicating the DFT simulation cells five times along both in-plane directions, resulting in cells containing 600--2000 atoms. In these larger cells, Ni atoms were substituted randomly not only within the GB plane but also in the adjacent layers, either by a single species (Cr or Fe) or simultaneously by both species to obtain a ternary composition, as described in Secs.~\ref{subsec:config_effect} and \ref{subsec:ternary}, respectively.

\subsection{Computational details}\label{subsec:comp_detail}

Electronic structure calculations were performed using the Vienna \textit{ab initio} simulation package (VASP) \cite{kresseEfficientIterativeSchemes1996}. The interactions between valence electrons and ionic cores were treated using the projector-augmented wave method \cite{blochlProjectorAugmentedwaveMethod1994,kresseUltrasoftPseudopotentialsProjector1999} within the generalized gradient approximation (GGA), employing the Perdew--Burke--Ernzerhof (PBE) exchange-correlation functional. The Ni $3d$ and $4s$, Fe $3d$ and $4s$, and Cr $3p$, $3d$ and $4s$ electrons were treated as valence states. A plane-wave energy cutoff of 600~eV was used. The Brillouin zone was sampled using Monkhorst--Pack \cite{monkhorstSPECIALPOINTSBRILLOUINZONE1976} $k$-point meshes chosen to maintain a reciprocal-space sampling density equivalent to that of a $12 \times 12 \times 12$ mesh for the fcc Ni unit cell. Gaussian smearing with a width of 0.1~eV was applied for the Brillouin-zone integrations \cite{elsasserDensityfunctionalEnergiesForces1994}. The electronic self-consistency loop was converged to $1 \times 10^{-6}$~eV, and ionic relaxations were continued until all interatomic forces were below $1 \times 10^{-2}$~eV/\AA{}.

Spin polarization was used to account for local magnetic moments. As for bulk fcc Ni, a ferromagnetic (FM) ordering was found to be more stable than antiferromagnetic (AFM) ordering for all the considered pure Ni GBs. For the Cr@GB and Fe@GB structures, several configurations with the local magnetic moments oriented parallel or antiparallel to the Ni moments were considered, and the lowest-energy configuration found in each case was selected for further investigation. The results are described in detail in App.~\ref{app:mag_moms}.

We introduced the separation above the GB plane by shifting all atoms on one side of the plane upward and elongating the simulation cell by the same amount. Energies were calculated at the following separations: $\{-0.2, -0.1, 0, 0.1, 0.2, 0.4, 0.5, 0.6, 0.7, 0.8, 1, 6\}$~\AA{}, generating energy--separation data. The separations were chosen to thoroughly sample both the region around the equilibrium distance and the region where the maximum traction occurs. Equation~\eqref{eq:uber} is fitted to these data points, and the work of separation $W_{\mathrm{sep}}$ is extracted. The maximum traction $\sigma_{\max}$ is subsequently obtained from Eq.~\eqref{eq:uber_derivative}.

To assess the performance of the GRACE foundation models, we performed the same RBS calculations, as done with DFT. Since the models may have different equilibrium lattice constants for pure nickel due to training, we proceeded similarly to the DFT approach: The lattice parameters in the $xy$ plane were fixed according to the model's equilibrium bulk lattice parameter, and the cell and atoms were relaxed along the $z$ axis prior to the RBS calculations until the relative energy change between two consecutive iterations in LAMMPS \cite{thompsonLAMMPSFlexibleSimulation2022} was below $10^{-8}$. Subsequently, the energy was calculated for the same separations as described above, and Eq.~\eqref{eq:uber} was fitted to the energy--separation data. The resulting values for the work of separation and maximum traction were compared with the DFT values using a combined metric that quantifies the mean relative error (MRE) averaged over both target quantities,

\begin{equation}
\begin{split}
\overline{\mathrm{MRE}}
= \frac{1}{2N}\sum_{i=1}^{N}
\Bigg[
& \frac{
    \left|W_{\mathrm{sep},i}^{\mathrm{GRACE}}
    - W_{\mathrm{sep},i}^{\mathrm{DFT}}\right|
}{
    W_{\mathrm{sep},i}^{\mathrm{DFT}}
}
\\
& + \frac{
    \left|\sigma_{\max,i}^{\mathrm{GRACE}}
    - \sigma_{\max,i}^{\mathrm{DFT}}\right|
}{
    \sigma_{\max,i}^{\mathrm{DFT}}
}
\Bigg],
\end{split}
\label{eq:mlip_metric}
\end{equation}
where $N$ is the number of RBS calculations, which in this work is 24. Normalizing by the DFT reference value makes both error terms dimensionless, so work of separation and maximum traction contribute equally to the combined score.

\section{Results and Discussion}\label{sec:results}


\subsection{Structural and energetic properties of Ni grain boundaries}
\label{subsec:gbwidening}

Prior to the RBS calculations, the Ni grain boundaries were structurally relaxed and subsequently characterized in terms of their geometric widening and grain boundary energy. Atomic positions were allowed to move along $z$, i.e., along the GB normal, and the $c$-lattice parameter was adjusted accordingly, while the in-plane Ni bulk lattice constant was kept fixed. This approximates the situation in which the two grains are large enough to recover bulk-like values away from the interface.

We define the geometric GB excess volume as

\begin{equation}
    \delta_{\mathrm{GB}}
    = \frac{
        V_{\mathrm{GB}}
        - N_{\mathrm{GB}} V_{\mathrm{bulk}}^{(\mathrm{at})}
    }{2A},
    \label{eq:gb_widening}
\end{equation}
where $V_{\mathrm{GB}}$ is the total volume of the relaxed grain boundary supercell, $N_{\mathrm{GB}}$ is the number of atoms in that supercell, $V_{\mathrm{bulk}}^{(\mathrm{at})}$ is the reference volume per atom of the undisturbed fcc-Ni bulk crystal, $A$ is the cross-sectional area of the grain boundary plane, and the factor of~$2$ accounts for the two equivalent grain boundaries contained in the periodic slab cell.
Table~\ref{tab:gb_energy_comparison} lists the resulting excess volumes and GB energies for the six pure Ni structures, together with GB energies from the literature. The calculated GB energies are generally in good agreement with the DFT literature values. The deviations may partly arise from differences in structural relaxation procedures. For example, in Refs.~\cite{beanOriginDifferencesExcess2016,zhengGrainBoundaryProperties2020}, all atomic positions and lattice parameters of the simulation cell were relaxed.

\begin{table}[b]
\centering
\caption{Grain boundary energies $\gamma_{\mathrm{GB}}$ compared to DFT values from the literature, together with the excess volumes $\delta_{\mathrm{GB}}$ for the investigated Ni GB structures.}
\begin{tabular}{lccc}
\toprule
\multirow{2}{*}{Grain boundary} & \multicolumn{2}{c}{$\gamma_{\mathrm{GB}}$ (J/m$^2$)} & \multirow{2}{*}{$\delta_{\mathrm{GB}}$ (\AA{})} \\
\cmidrule(lr){2-3}
 & This work & Ref. & \\
\midrule
$\Sigma3~(111)~[\bar{1}10]$ & 0.07 & 0.06 \cite{sestakExtraordinaryResponseHCharged2020} & 0.01 \\
$\Sigma3~(112)~[\bar{1}10]$ & 0.94 & 0.84 \cite{beanOriginDifferencesExcess2016} & 0.15 \\
$\Sigma5~(012)~[100]$ & 1.28 & 1.31 \cite{mazalovaEffectVacanciesGrain2020} & 0.34 \\
$\Sigma5~(013)~[100]$ & 1.25 & 1.26 \cite{zhengGrainBoundaryProperties2020} & 0.30 \\
$\Sigma9~(221)~[\bar{1}10]$ & 1.18 & 1.16 \cite{zhengGrainBoundaryProperties2020} & 0.31 \\
$\Sigma11~(113)~[\bar{1}10]$ & 0.44 & 0.43 \cite{sestakExtraordinaryResponseHCharged2020} & 0.13 \\
\bottomrule
\end{tabular}
\label{tab:gb_energy_comparison}
\end{table}

\begin{figure}
\includegraphics[width=\columnwidth]{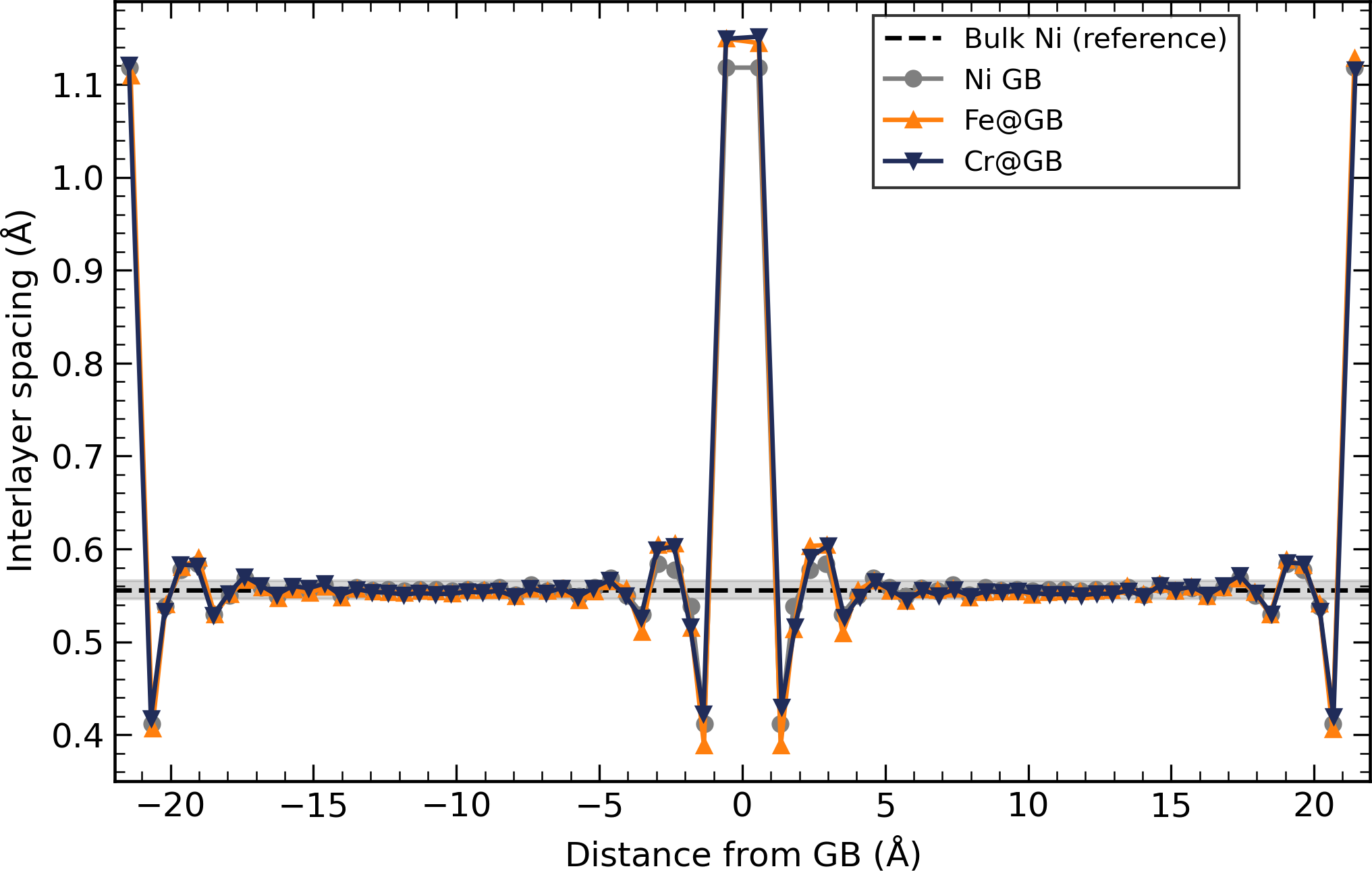}
\caption{Interlayer spacing along the grain boundary normal for the $\Sigma5~(013)~[100]$ GB, comparing pure Ni (Ni GB) with Cr- and Fe-substituted grain boundaries (Cr@GB, Fe@GB). The dashed horizontal line marks the bulk Ni interlayer spacing as reference. Distances are reported relative to the GB plane, located at fractional $z = 0.5$. The region around $\pm$0.01~\AA{} of the bulk interlayer distance is shaded in grey.}
\label{fig:gb_widening}
\end{figure}

Figure~\ref{fig:gb_widening} shows the interlayer distances in the $\Sigma5~(013)~[100]$ GB for pure Ni, as well as for structures containing Fe or Cr. While Cr and Fe substitutions do have an effect, the main difference compared to the interlayer distances in the bulk is attributable to the geometric properties of the grain boundaries. Between the two GB planes of the simulation cell, the interlayer spacings can be seen to approach the bulk value (within 0.01~\AA{}), supporting the assumption stated above. This behavior was verified for all six GB structures investigated in this work. 

\subsection{Chemical effect of substitutes at the GB}\label{subsec:dft_res}

\begin{figure}
\includegraphics[width=0.9\columnwidth]{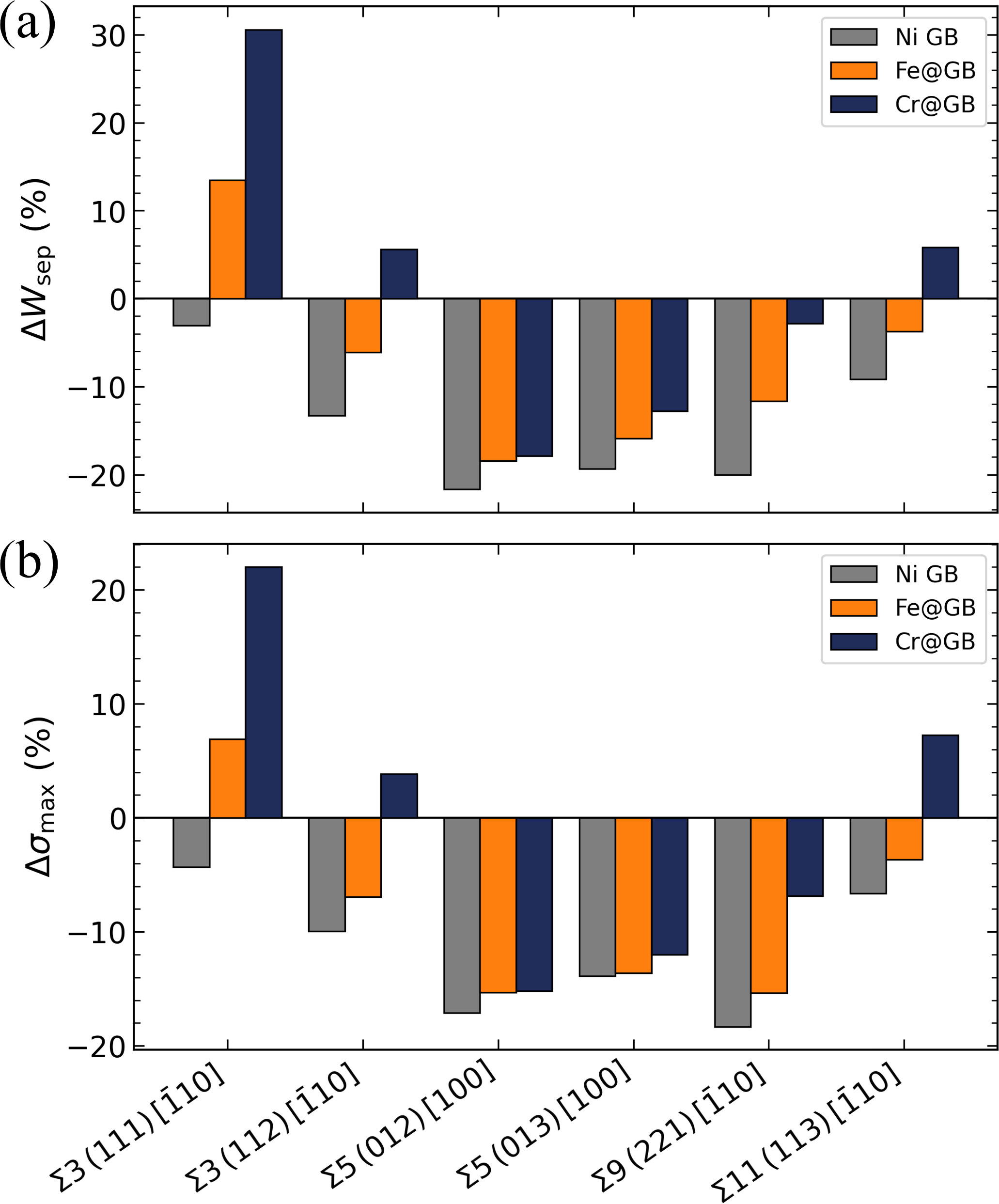}
\caption{Relative change in work of separation $W_{\mathrm{sep}}$ (a) and maximum traction $\sigma_{\max}$ (b) for the pure Ni GB and the Cr@GB and Fe@GB structures, relative to the pure Ni bulk value.}
\label{fig:chem_effect}
\end{figure}

For each of the six GBs, four RBS calculations were performed: one reference calculation on the corresponding bulk cell (where the separation is applied normal to the plane, which is the GB plane in the GB structures), and three calculations on the GB structures, namely the pure Ni GB and the two substituted configurations, Cr@GB and Fe@GB. The absolute values for all structures are reported in Table~\ref{tab:uber_species_vs_bulk}, and Fig.~\ref{fig:chem_effect} shows the relative deviation of the work of separation $W_{\mathrm{sep}}$ and the maximum traction $\sigma_{\max}$ with respect to bulk Ni. These relative deviations are defined as
\begin{equation}
    \Delta W_{\mathrm{sep}}
    = \frac{
        W_{\mathrm{sep}}^{\mathrm{GB}}
        - W_{\mathrm{sep}}^{\mathrm{bulk}}
    }{W_{\mathrm{sep}}^{\mathrm{bulk}}},
    \label{eq:rel_deviation_Wsep}
\end{equation}
and
\begin{equation}
    \Delta \sigma_{\max}
    = \frac{
        \sigma_{\max}^{\mathrm{GB}}
        - \sigma_{\max}^{\mathrm{bulk}}
    }{\sigma_{\max}^{\mathrm{bulk}}},
    \label{eq:rel_deviation_sig}
\end{equation}
where $W_{\mathrm{sep}}^{\mathrm{GB}}$ and $\sigma_{\max}^{\mathrm{GB}}$ denote the values for the pure Ni GB, Cr@GB, or Fe@GB configuration, and $W_{\mathrm{sep}}^{\mathrm{bulk}}$ and $\sigma_{\max}^{\mathrm{bulk}}$ are the corresponding bulk reference values. 

For all tested GBs, the pure Ni GBs exhibit lower values for $W_{\mathrm{sep}}$ and $\sigma_{\max}$ than the corresponding bulk structures. This can be explained by the non-equilibrium atomic distances and different coordination environments compared to bulk Ni, resulting from the local geometries at the GBs \cite{MaterialsInterfaces,suttonInterfacesCrystallineMaterials2006}. However, the quantitative reduction varies between grain boundaries. While the values for the $\Sigma3~(111)~[\bar{1}10]$ GB, which has a low GB energy, are close to the bulk values, for grain boundaries with higher GB energies, the values are much lower compared to the bulk (up to approx.\ $-20~\%$ for $\Sigma5~(012)~[100]$ or $\Sigma9~(221)~[\bar{1}10]$). 

The work of separation values agree well with the values for unrelaxed surfaces reported by \v{S}est\'{a}k \textit{et al.}~\cite{sestakExtraordinaryResponseHCharged2020} for the $\Sigma3~(111)~[\bar{1}10]$ and $\Sigma11~(113)~[\bar{1}10]$ GBs (3.78 and 4.35~J/m$^2$). For the $\Sigma5~(012)~[100]$ GB, \v{C}ern\'{y} \textit{et al.}~\cite{cernyInitioTensileTests2016} reported 3.94~J/m$^2$ and 25.3~GPa for $W_{\mathrm{sep}}$ and $\sigma_{\max}$, respectively, which match our values almost exactly.

Chen and Dongare~\cite{chenRoleGrainBoundary2017a} reported maximum tensile stresses of 28.06, 24.89, 23.55, and 27.51~GPa for the pristine $\Sigma3~(111)~[\bar{1}10]$, $\Sigma5~(012)~[100]$, $\Sigma5~(013)~[100]$, and $\Sigma11~(113)~[\bar{1}10]$ GBs, respectively. For $\Sigma3~(111)~[\bar{1}10]$, $\Sigma5~(012)~[100]$, and $\Sigma11~(113)~[\bar{1}10]$, these agree with our values of 29.07, 25.39, and 28.40~GPa within the variation expected due to differences in the computational details. For $\Sigma5~(013)~[100]$, however, our value of 26.64~GPa is substantially higher than their 23.55~GPa. We attribute this to the GB structure: in our construction, two closely spaced Ni atoms were merged into one site (Sec.~\ref{subsec:strucs}), whereas Chen and Dongare retain both, resulting in a significantly higher GB energy of 1.91~J/m$^2$ compared to our 1.25~J/m$^2$.

\begin{table*}[htbp]
\centering
\caption{Work of separation $W_{\mathrm{sep}}$ and maximum traction $\sigma_{\max}$ for the considered grain boundary structures (pure Ni GB and with Fe or Cr as substitutional element at the GB plane). Values for the rigid body separation of a Ni bulk crystal along the same lattice plane are given as reference.}
\label{tab:uber_species_vs_bulk}
\begin{tabular}{lcccccccc}
\toprule
 & \multicolumn{2}{c}{Ni bulk crystal} & \multicolumn{2}{c}{Ni GB} & \multicolumn{2}{c}{Cr@GB} & \multicolumn{2}{c}{Fe@GB} \\
\cmidrule(lr){2-3} \cmidrule(lr){4-5} \cmidrule(lr){6-7} \cmidrule(lr){8-9}
GB & $W_{\mathrm{sep}}$ (J/m$^2$) & $\sigma_{\max}$ (GPa) & $W_{\mathrm{sep}}$ (J/m$^2$) & $\sigma_{\max}$ (GPa) & $W_{\mathrm{sep}}$ (J/m$^2$) & $\sigma_{\max}$ (GPa) & $W_{\mathrm{sep}}$ (J/m$^2$) & $\sigma_{\max}$ (GPa) \\
\midrule
$\Sigma3~(111)~[\bar{1}10]$ & 3.91 & 30.38 & 3.79 & 29.07 & 5.11 & 37.07 & 4.44 & 32.48 \\
$\Sigma3~(112)~[\bar{1}10]$ & 4.63 & 30.39 & 4.02 & 27.36 & 4.89 & 31.56 & 4.35 & 28.28 \\
$\Sigma5~(012)~[100]$ & 5.03 & 30.63 & 3.94 & 25.39 & 4.13 & 25.98 & 4.11 & 25.94 \\
$\Sigma5~(013)~[100]$ & 5.05 & 30.94 & 4.07 & 26.64 & 4.40 & 27.24 & 4.24 & 26.72 \\
$\Sigma9~(221)~[\bar{1}10]$ & 4.45 & 30.20 & 3.56 & 24.66 & 4.32 & 28.13 & 3.93 & 25.55 \\
$\Sigma11~(113)~[\bar{1}10]$ & 4.87 & 30.41 & 4.42 & 28.40 & 5.15 & 32.62 & 4.69 & 29.29 \\
\bottomrule
\end{tabular}
\end{table*}

There is an approximately linear trend between GB energy and the reduction in $W_{\mathrm{sep}}$ and $\sigma_{\max}$ (with a Pearson correlation coefficient \cite{pearsonVIINoteRegression1997} between the magnitude of the reduction in $W_{\mathrm{sep}}$ and GB energy of $r = 0.98$, and between the magnitude of the reduction in $\sigma_{\max}$ and GB energy of $r = 0.93$), where deviations from a perfect correlation can be explained by the fact that the RBS calculations primarily capture the modified local bonding environments along the direction of separation, while the grain boundary energy captures the spatially averaged effect of the local geometry across the interface.

Replacing Ni atoms with Fe or Cr in the grain boundary plane increases both the work of separation and the maximum traction. This is consistent with thermodynamic calculations for the ternary fcc system, which show that mixed Fe--Ni and Cr--Ni bonds are more energetically favorable than the corresponding pure bonds \cite{wrobelPhaseStabilityTernary2015}. The effective interaction between Cr and Ni in ternary Ni--Fe--Cr alloys is significantly stronger than that between Fe and Ni, which is consistent with the observation that the increase in the quantities in Fig.~\ref{fig:chem_effect} caused by Cr is greater than that caused by Fe. However, the extent of the increase varies strongly and does not follow any easily discernible trend, which can be explained by the different local geometric environments of the various grain boundaries.


\subsection{GRACE foundation models}\label{subsec:mlip_parity}

In order to overcome the size limitations implied by the computational cost of DFT simulations, we use the DFT data described in the previous section as a reference to evaluate the performance of various MLIPs and subsequently select one for larger and more numerous simulations.
The comparison to DFT constitutes a particularly stringent test for the MLIPs, as they require an accurate description not only of bulk properties, but also of surface energies (at large separations), the local structural motifs at the grain boundaries, and the stretching of bonds across the full range of separations. We evaluated 26 GRACE foundation models against the DFT reference data using the combined mean relative error $\overline{\mathrm{MRE}}$ defined in Eq.~\eqref{eq:mlip_metric}. Most of the models yield similar results, with 17 of the 26 reaching $\overline{\mathrm{MRE}} < 3~\%$. The five models with the lowest $\overline{\mathrm{MRE}}$ (1.17--1.79~\%) differ in architecture and size, but share the same training data: they are all trained on OMAT \cite{barros-luqueOpenMaterials20242026}, Alexandria \cite{cavignacAIDrivenExpansionApplication2026}, and Materials Project (MP) \cite{jainCommentaryMaterialsProject2013} datasets (-ft-AM). The GRACE-2L-OMAT-large-ft-AM model achieved the lowest $\overline{\mathrm{MRE}}$ value and was therefore chosen as the model for further investigations. The values of MAE and $\overline{\mathrm{MRE}}$ for all models are summarized in App.~\ref{app:mlip_scores}.

\begin{figure}
\includegraphics[width=\columnwidth]{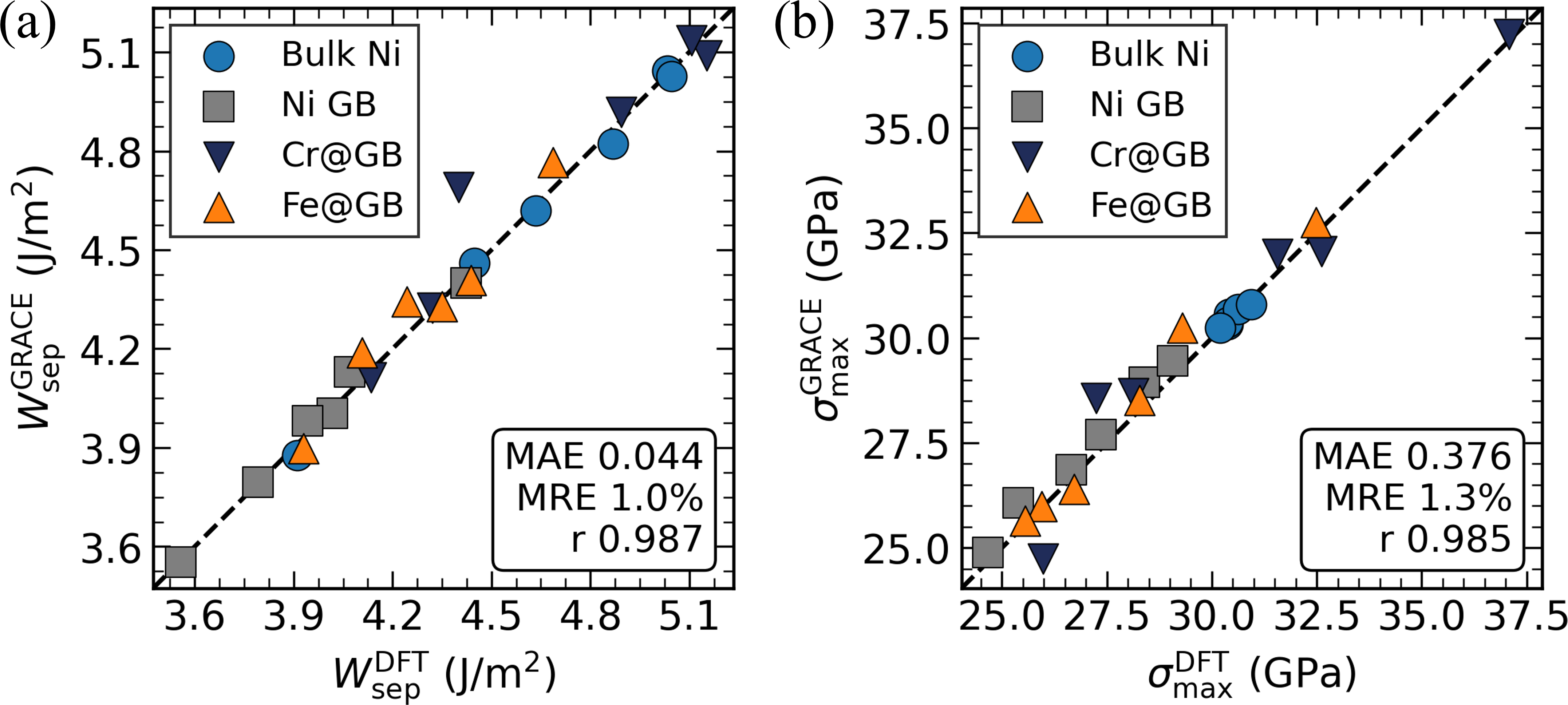}
\caption{Parity plots comparing the work of separation (a) and maximum traction (b), obtained with the GRACE-2L-OMAT-large-ft-AM foundation model against DFT reference values. The mean absolute error (MAE), the mean relative error (MRE) and the Pearson correlation coefficient $r$ \cite{pearsonVIINoteRegression1997} are reported in the legends of each quantity.}
\label{fig:parity}
\end{figure}

The comparison between the values obtained by DFT and the GRACE-2L-OMAT-large-ft-AM model is presented in Fig.~\ref{fig:parity}. The figure shows that the model exhibits excellent agreement with the DFT data for the Ni bulk and Ni GB systems (mean absolute error ($\mathrm{MAE}$) of $W_{\mathrm{sep}}$: $0.023$~J/m$^2$ and $0.024$~J/m$^2$, respectively). For the Fe@GB structures, $\mathrm{MAE}(W_{\mathrm{sep}}) = 0.057$~J/m$^2$. The Cr@GB structures show the largest deviations with $\mathrm{MAE}(W_{\mathrm{sep}}) = 0.072$~J/m$^2$. This is likely due, in part, to the magnetic degrees of freedom, which are not taken into account explicitly in the MLIP. For Fe@GB, the FM arrangements were the most energetically favorable for all systems (see App.~\ref{app:mag_moms}), essentially reducing this degree of freedom. In contrast, the preferred magnetic arrangement for Cr@GB depends on the GB structure. The largest deviation in $W_{\mathrm{sep}}$ occurs for the $\Sigma5~(013)~[100]$ Cr@GB structure, where the two Cr atoms carry large magnetic moments of opposite sign ($-1.81$ and $+2.56\,\mu_\mathrm{B}$). The absence of explicit magnetic degrees of freedom in the MLIP likely contribute to the larger deviations observed for Cr@GB.

\subsection{Configurational effects of binary substitution at the GB}\label{subsec:config_effect}

To achieve a more realistic description of the material, larger simulation cells are needed in which, for example, grain boundaries with larger structural motifs can be realized, extending all the way to general grain boundaries \cite{itoMachineLearningInteratomic2024a}. Here, we aim to use the GRACE MLIP to describe the alloying effect of Cr and Fe at the grain boundaries more realistically. To do this, we use simulation cells that are five times larger within the grain boundary plane in both directions, reaching sizes of 600--2000 atoms. 

\begin{figure}
\includegraphics[width=\columnwidth]{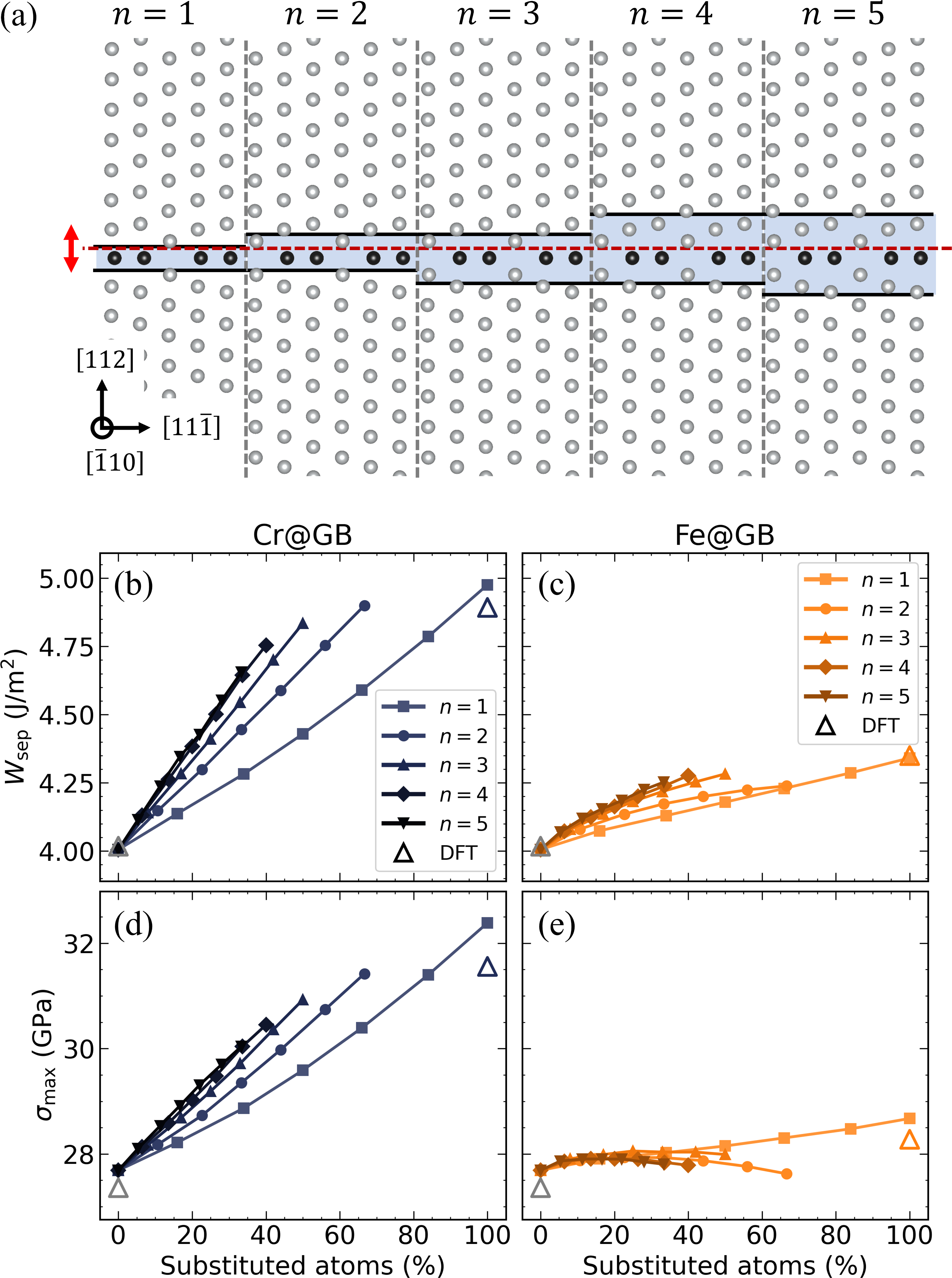}
\caption{Effect of alloying species on GB cohesion. (a): The $\Sigma3~(112)~[\bar{1}10]$ GB structure, with the GB plane indicated by darker shaded atoms. From left to right, the increasing number of Ni layers $n$ considered for substitution is highlighted. The horizontal dashed line indicates the plane where the separation is introduced. (b) and (c): Mean work of separation obtained from UBER fits to RBS calculations performed with the GRACE MLIP for different concentrations of the substitute elements Cr and Fe, respectively, and different numbers of layers $n$. (d) and (e): Mean maximum traction derived from the same RBS calculations.}
\label{fig:config_effect}
\end{figure}

The MLIP not only allows us to calculate many different arrangements of the alloying elements but also, due to the larger simulation cells, enables finer gradations in element concentrations. In the DFT analysis, we considered only substitutions within the grain boundary plane to keep the number of possible structures and the associated magnetic arrangements manageable. In the following, we extend the analysis beyond substitution solely within the grain boundary plane by also substituting atoms in the layers above and below it, while the separation during the RBS calculations is still performed directly above the GB plane. We first examined how many alloyed layers need to be considered before the results no longer change significantly. Distributing the substitute atoms over several layers, rather than placing them exclusively in the GB plane, gives a more realistic picture and approaches a randomly mixed alloy without preferential segregation to the GB. This is shown exemplarily in Fig.~\ref{fig:config_effect} for the $\Sigma3~(112)~[\bar{1}10]$ GB. Subfigure (a) in Fig.~\ref{fig:config_effect} shows the increasing number of layers considered for substitution.

\begin{figure*}[htp]
\includegraphics[width=\textwidth]{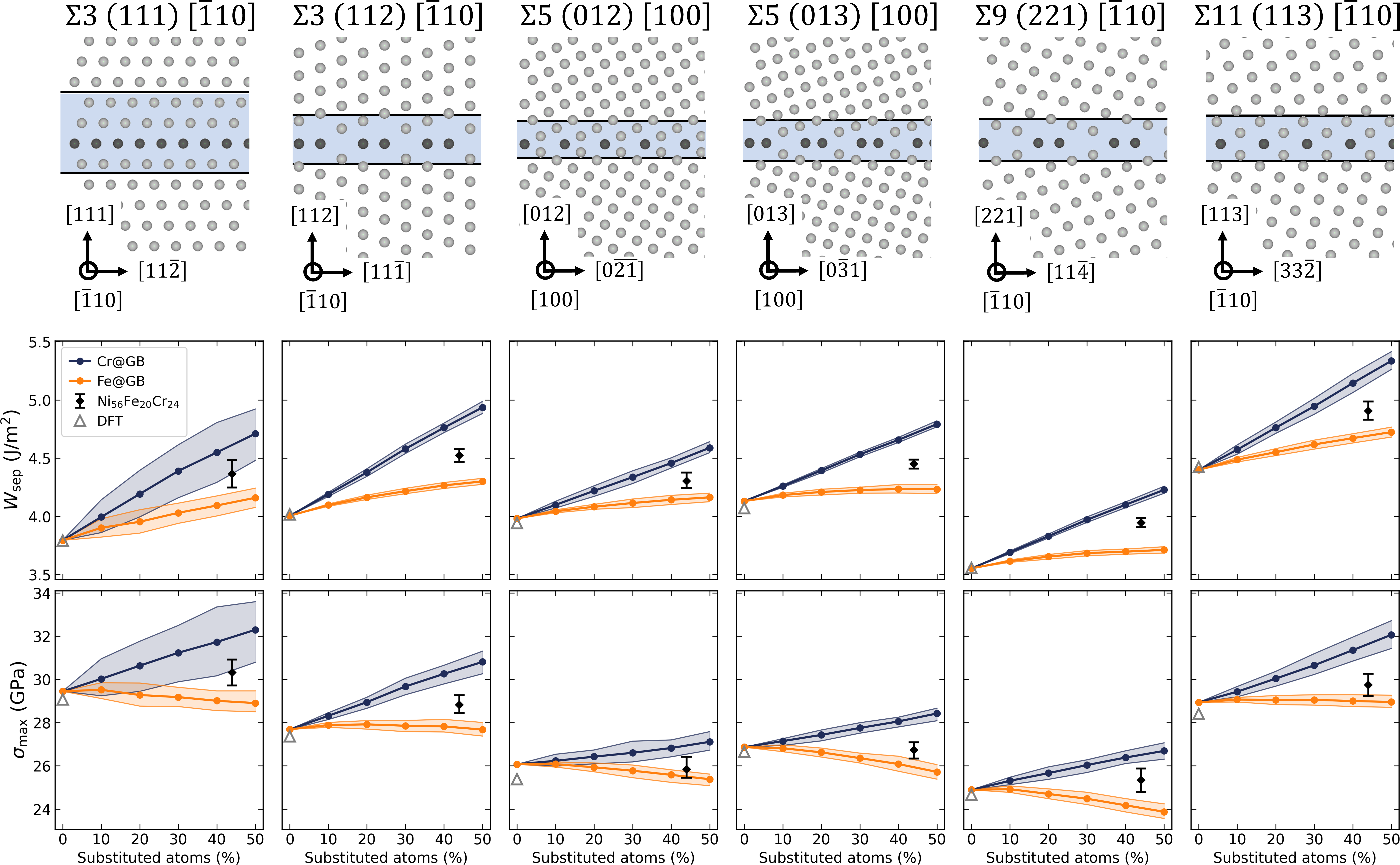}
\caption{Top: Layers considered for the substitution of Ni atoms with Cr or Fe in the six grain boundaries studied. Only atoms located entirely within the shaded region were considered for substitution. The darker-shaded atoms are located within the grain boundary plane. Middle: Work of separation $W_{\mathrm{sep}}$ obtained from MLIP-based RBS calculations. Circles show the mean values over 100 random binary substitution configurations per concentration, and shaded bands span the minimum and maximum values. Black diamonds represent the ternary composition $\mathrm{Ni}_{56}\mathrm{Fe}_{20}\mathrm{Cr}_{24}$, with bars indicating the minimum and maximum values. Bottom: Maximum traction $\sigma_{\max}$, plotted in the same manner.}
\label{fig:gb_strucs_subs_range}
\end{figure*}

For each concentration and number of layers considered, 50 random arrangements of Cr or Fe were generated. This means that out of all Ni atoms located within the considered layers, the atoms to be substituted by Cr or Fe were selected randomly. For each number of layers, the maximum number of substituted atoms was limited to the number corresponding to a concentration of 100~\% when substitution is restricted to the GB plane alone. Higher concentrations for a larger number of layers would otherwise correspond to an overall higher concentration of Cr or Fe atoms. This is particularly relevant because the in-plane lattice parameters are kept fixed to simulate bulk Ni between the GBs, such that a high concentration of substitute atoms could give rise to additional, confounding effects. Using the GRACE potential, each of these simulation cells was relaxed along the GB normal (cf.\ also Sec.~\ref{subsec:gbwidening}), RBS calculations were performed, and Eq.~\eqref{eq:uber} was fitted to these data.
Figure~\ref{fig:config_effect} (b) and (c) show the mean work of separation $W_{\mathrm{sep}}$ as a function of concentration for the different numbers of layers considered, for the Cr@GB and Fe@GB systems, respectively. Subfigures (d) and (e) in Fig.~\ref{fig:config_effect} show the values for $\sigma_{\max}$ in the same manner.

For the case where only substitution within the GB plane was considered, both $W_{\mathrm{sep}}$ and $\sigma_{\max}$ increase linearly with increasing concentration. The deviations from the DFT data are the same ones visible in Fig.~\ref{fig:parity}, i.e., they reflect the differences between DFT and the chosen GRACE potential. This linear trend is consistent with the interpretation that, as the concentration increases, more Cr--Ni and Fe--Ni bonds are broken during separation, as discussed in Sec.~\ref{subsec:dft_res}.

In contrast, substituting Ni atoms in additional layers leads to two qualitatively different phenomena:
For the Cr@GB systems (subfigures (b) and (d) in Fig.~\ref{fig:config_effect}), we continue to observe an approximately linear trend in both quantities, with the slope increasing as more layers are included. This suggests that Cr--Cr bonds also contribute to strengthening the interfacial cohesion. This increase in slope appears to converge for four layers, with only minor deviations in both $W_{\mathrm{sep}}$ and $\sigma_{\max}$ when a fifth layer is substituted. This is plausible, since substitutions in layers further away from the GB plane are expected to have a progressively smaller effect on the interface cohesion.
The Fe@GB systems exhibit different effects. While Fe--Ni bonds increase $W_{\mathrm{sep}}$ and $\sigma_{\max}$ (see Fig.~\ref{fig:config_effect} (c) and (e)), albeit to a lesser extent than Cr--Ni bonds, Fe--Fe bonds being stretched across the separation plane lead to different behavior. As with the Cr@GB systems, $W_{\mathrm{sep}}$ increases with higher concentrations, though not as linearly as when substitution is confined to the GB plane alone. The maximum traction, on the other hand, only increases for small Fe concentrations when more layers are considered, then remains approximately constant and even decreases at high Fe concentrations. Within the UBER description, the simultaneous increase in $W_{\mathrm{sep}}$ and decrease in $\sigma_{\max}$ implies that the characteristic length $L$ increases with Fe concentration by a larger relative amount than $W_{\mathrm{sep}}$, see Eq.~\eqref{eq:uber_max_traction}. The resulting broader traction--separation curve therefore has a lower peak traction despite the greater total energy required for separation. Again, the results vary only slightly between four and five layers considered, and we therefore chose to investigate substitution for this GB structure using four layers.

The same procedure was applied to the other GBs studied here, and the layers considered for substitution are marked in the top row of Fig.~\ref{fig:gb_strucs_subs_range}. The middle and bottom subfigures show the resulting values for the work of separation $W_{\mathrm{sep}}$ and the maximum traction $\sigma_{\max}$, respectively. We calculated both quantities for all GBs up to concentrations of 50~\%. Beyond this point, as described above, confounding effects arising from high overall concentrations of Cr or Fe in the simulation cells are expected. The results shown are the minimum, maximum, and mean values for 100 random arrangements. In order to check the convergence of the mean values, 10 additional random arrangements were generated for each GB, substitutional element, and concentration, and the corresponding RBS calculations were performed. In all cases, the mean values changed by at most 0.1~\%. While a larger number of calculations might reveal even lower minimum or higher maximum values, the small change in the mean values indicates that the mean provides a statistically reliable estimate of the expected value, under the assumption that the substitutional elements are randomly distributed. 

For the $\Sigma3~(111)~[\bar{1}10]$, $\Sigma5~(012)~[100]$, $\Sigma5~(013)~[100]$, and $\Sigma9~(221)~[\bar{1}10]$ GBs, the trends with increasing Cr or Fe concentration follow the same qualitative behavior we have already discussed for $\Sigma3~(112)~[\bar{1}10]$. In the Cr@GB systems, both $W_{\mathrm{sep}}$ and $\sigma_{\max}$ increase approximately linearly with Cr concentration, while in the Fe@GB systems $W_{\mathrm{sep}}$ follows the same trend (although for $\Sigma5~(013)~[100]$ it is nearly constant), but $\sigma_{\max}$ decreases with increasing Fe content. For $\Sigma11~(113)~[\bar{1}10]$, the same trends hold, with one exception: here, $\sigma_{\max}$ in the Fe@GB systems does not decrease with increasing Fe content but instead remains essentially constant across the entire concentration range. All absolute values are reported in App.~\ref{app:config_effect_all}.

\subsection{GB cohesion for a ternary composition}\label{subsec:ternary}

While the binary substitutions allow the effects of Fe and Cr to be examined separately, technologically relevant Ni-based alloys contain both elements simultaneously. As a first step toward such multicomponent systems, we therefore extended the calculations to a ternary composition of $\mathrm{Ni}_{56}\mathrm{Fe}_{20}\mathrm{Cr}_{24}$ (at.\%), inspired by the principal constituents of Alloy~718 \cite{zhaoMultiscaleModelingMethod2022a}. Within the substitution regions shown in Fig.~\ref{fig:gb_strucs_subs_range}, Ni atoms were randomly replaced by Fe and Cr to obtain the prescribed composition, corresponding to a total substitution concentration of 44~\%. For each GB structure, 100 random configurations were generated. The resulting mean values of $W_{\mathrm{sep}}$ and $\sigma_{\max}$ are included in Fig.~\ref{fig:gb_strucs_subs_range}, and the numerical values are reported in Table~\ref{tab:ternary_stats}.

\begin{table}[htbp]
\centering
\caption{Work of separation $W_{\mathrm{sep}}$ and maximum traction $\sigma_{\max}$ for Ni grain boundaries with Ni atoms randomly replaced by Fe and Cr within the substitution regions shown in Fig.~\ref{fig:gb_strucs_subs_range}, yielding a local composition of $\mathrm{Ni}_{56}\mathrm{Fe}_{20}\mathrm{Cr}_{24}$, inspired by the principal constituents of Alloy~718. Values are reported as mean $\pm$ standard deviation over 100 configurations.}
\label{tab:ternary_stats}
\begin{tabular}{lcc}
\toprule
GB & $W_{\mathrm{sep}}$ (J/m$^2$) & $\sigma_{\max}$ (GPa) \\
\midrule
$\Sigma 3~(111)~[\bar{1}10]$ & $4.37 \pm 0.06$ & $30.32 \pm 0.30$ \\
$\Sigma 3~(112)~[\bar{1}10]$ & $4.52 \pm 0.02$ & $28.82 \pm 0.18$ \\
$\Sigma 5~(012)~[100]$ & $4.30 \pm 0.02$ & $25.84 \pm 0.19$ \\
$\Sigma 5~(013)~[100]$ & $4.45 \pm 0.02$ & $26.73 \pm 0.16$ \\
$\Sigma 9~(221)~[\bar{1}10]$ & $3.95 \pm 0.02$ & $25.33 \pm 0.19$ \\
$\Sigma 11~(113)~[\bar{1}10]$ & $4.91 \pm 0.04$ & $29.74 \pm 0.21$ \\
\bottomrule
\end{tabular}
\end{table}

For both quantities, the ternary results lie between the corresponding binary Cr@GB and Fe@GB values at the same total substitution concentration, consistent with a balance between the individual effects of the two substitutes. However, this intermediate response is not necessarily captured by a linear interpolation of the binary results. For the $\Sigma 5\,(012)\,[100]$ and $\Sigma 5\,(013)\,[100]$ GBs, for example, the ternary $\sigma_{\max}$ lies closer to the Fe@GB curve despite the comparable Fe and Cr fractions, suggesting a disproportionate influence of Fe on the maximum traction. The binary calculations therefore provide a useful basis for interpreting the ternary response, but explicit ternary calculations are needed to assess the combined solute effects quantitatively.

Taken together, the GB-dependent binary trends and the effects observed in the ternary model highlight the need to resolve both boundary structure and local alloy composition when assessing GB cohesion. A systematic investigation of these dependencies across many GBs, alloying elements, and configurations would be impractical with DFT alone at the scale considered here. It is precisely this limitation that underlines the value of the GRACE MLIP. Beyond identifying qualitative trends, its good quantitative agreement with DFT (see Sec.~\ref{subsec:mlip_parity}) enables the calculation of statistically well-founded values for $W_{\mathrm{sep}}$ and $\sigma_{\max}$. Accounting for Fe and Cr simultaneously, the ternary calculations extend the binary analysis to more realistic chemical environments and represent a first step toward multicomponent alloys. The resulting cohesive properties can be used, for example, to parameterize cohesive-zone models in FEM simulations \cite{mollerFractureInitioForcebased2018}, providing a route toward atomistically informed multiscale modeling of such alloys.

\section{Summary}\label{sec:conc}

In this work, we investigated the influence of Cr and Fe alloying on the cohesive properties of Ni grain boundaries by combining DFT-based rigid-body separation calculations with a machine-learning interatomic potential based on the GRACE framework. DFT calculations across six symmetric-tilt GBs showed that substitution of Cr or Fe at the GB plane consistently increases both the work of separation and the maximum traction relative to pure Ni GBs. However, the magnitude of this effect varies substantially between GB structures, reflecting the distinct local geometric environments at each interface, and precludes a simple, transferable trend across GB types.

To extend this analysis beyond the size and configurational limitations of DFT, we compared 26 GRACE foundation models and identified one achieving good quantitative agreement with the DFT reference data for both $W_{\mathrm{sep}}$ and $\sigma_{\max}$. Using this MLIP, we performed large-scale RBS calculations for several thousand random alloying configurations at fine concentration resolution, a level of configurational sampling beyond what is feasible with DFT. This revealed distinct concentration-dependent trends for Cr and Fe, which arise from differences in their respective bonding contributions to GB cohesion. For a ternary $\mathrm{Ni}_{56}\mathrm{Fe}_{20}\mathrm{Cr}_{24}$ composition, both, $W_{\mathrm{sep}}$ and $\sigma_{\max}$ lie between the corresponding binary results, but the relative contributions of Fe and Cr vary with the GB structure. Thus, the binary results aid interpretation, while a quantitative description requires explicit multicomponent calculations.

These results demonstrate that DFT-validated MLIPs provide a practical and reliable route to capture configurational and chemical alloying effects at grain boundaries that would otherwise remain inaccessible due to the computational cost of first-principles methods. The statistically well-founded values for $W_{\mathrm{sep}}$ and $\sigma_{\max}$ obtained here, together with the general workflow presented, are readily transferable to other multicomponent alloy systems and provide a solid foundation for parameterizing continuum-scale cohesive-zone models of grain boundary fracture.

\begin{acknowledgments}
This work was funded by the German Research Foundation (DFG) under project number 526257118. Crystallographic drawings were created with the software VESTA \cite{mommaVESTA3Threedimensional2011a}.
\end{acknowledgments}

\bibliography{references}

\clearpage
\onecolumngrid
\appendix

\section{Magnetic ground states of Cr@GB and Fe@GB}
\label{app:mag_moms}

To test various magnetic configurations for the Cr@GB and Fe@GB structures, the Ni atoms were initialized in an FM configuration, and each GB atom (Cr or Fe) was assigned an initial magnetic moment of $\pm 1$ or $\pm 5\,\mu_{\mathrm{B}}$, where the sign denotes the orientation relative to the Ni magnetic moments. When more than one Cr or Fe atom was present at the GB, all combinations of $\pm$ orientations across these atoms were enumerated for each moment magnitude. In addition, a non-magnetic (NM) configuration was included for comparison.

Across all six GBs, the lowest-energy Fe@GB configurations found exhibit Fe moments aligned parallel to the Ni magnetization and, where applicable, to the other Fe moments. The local Fe moments range from approximately $2.5$ to $3.0\,\mu_\mathrm{B}$, and neighboring Ni moments remain essentially unperturbed ($\mu_\mathrm{Ni}\approx 0.58$--$0.66\,\mu_\mathrm{B}$). For Cr@GB, the preferred local magnetic configuration depends on the GB structure. The Cr moments are essentially quenched in $\Sigma3~(111)~[\bar{1}10]$, $\Sigma3~(112)~[\bar{1}10]$, and $\Sigma11~(113)~[\bar{1}10]$, with $|\mu_\mathrm{Cr}|<0.2\,\mu_\mathrm{B}$. For $\Sigma5~(013)~[100]$ and $\Sigma9~(221)~[\bar{1}10]$, the two Cr atoms carry oppositely oriented moments of unequal magnitude. For $\Sigma5~(012)~[100]$, the Cr moment is $-2.60\,\mu_\mathrm{B}$ and is antiparallel to the Ni magnetization. This Cr-induced quenching and frustration is also reflected in the neighboring Ni moments, which are locally suppressed to $0.13$--$0.30~\mu_\mathrm{B}$ in the Cr@GB structures, in stark contrast to the Fe@GB cases. These results are qualitatively consistent with the preference for antiparallel Cr alignment and parallel Fe alignment reported for bulk Ni--Fe--Cr alloys using magnetic cluster expansion \cite{suFirstprinciplesClusterExpansion2024}.

\begin{table}[htbp]
\centering
\caption{Local magnetic moments of the Cr and Fe substitute atoms in the lowest-energy configuration for each GB. The sign refers to the orientation of the moments with respect to the Ni atoms.}
\label{tab:mag_moments}
\begin{tabular}{lll}
\toprule
Grain boundary & Cr moments ($\mu_\mathrm{B}$) & Fe moments ($\mu_\mathrm{B}$) \\
\midrule
$\Sigma3~(111)~[\bar{1}10]$  & $+0.04$ & $+2.53$ \\
$\Sigma3~(112)~[\bar{1}10]$  & $+0.17,\ -0.01$ & $+2.59,\ +2.74$ \\
$\Sigma5~(012)~[100]$  & $-2.60$ & $+2.97$ \\
$\Sigma5~(013)~[100]$  & $-1.81,\ +2.56$ & $+2.75,\ +2.99$ \\
$\Sigma9~(221)~[\bar{1}10]$  & $-0.46,\ +0.74$ & $+2.64,\ +2.81$ \\
$\Sigma11~(113)~[\bar{1}10]$  & $-0.02$ & $+2.78$ \\
\bottomrule
\end{tabular}
\end{table}

\clearpage

\section{Comparison metrics for all GRACE models versus DFT data}
\label{app:mlip_scores}

Table~\ref{tab:uber_category_mae} reports the mean absolute error between the GRACE foundation models and the DFT data obtained from RBS calculations for the work of separation $W_{\mathrm{sep}}$ and the maximum traction $\sigma_{\max}$. Models are ordered by ascending combined mean relative error $\overline{\mathrm{MRE}}$ (Eq.~\eqref{eq:mlip_metric}), which is listed in the rightmost column and used to select the model for further calculations.

\begin{table}[htbp]
\centering
\caption{Mean absolute errors of the predicted work of separation $W_{\mathrm{sep}}$ and maximum traction $\sigma_{\max}$ relative to the DFT reference values, grouped by structural category. The final column gives the overall combined mean relative error $\overline{\mathrm{MRE}}$ (\%). Models are ordered by ascending $\overline{\mathrm{MRE}}$.}
\label{tab:uber_category_mae}
\begin{tabular}{lccccccccc}
\toprule
Model
& \multicolumn{2}{c}{Bulk Ni}
& \multicolumn{2}{c}{Ni GB}
& \multicolumn{2}{c}{Cr@GB}
& \multicolumn{2}{c}{Fe@GB}
& $\overline{\mathrm{MRE}}$ (\%) \\
\cmidrule(lr){2-3}
\cmidrule(lr){4-5}
\cmidrule(lr){6-7}
\cmidrule(lr){8-9}
& MAE & MAE & MAE & MAE & MAE & MAE & MAE & MAE & \\
& $W_{\mathrm{sep}}$ & $\sigma_{\max}$
& $W_{\mathrm{sep}}$ & $\sigma_{\max}$
& $W_{\mathrm{sep}}$ & $\sigma_{\max}$
& $W_{\mathrm{sep}}$ & $\sigma_{\max}$ & \\
& (J/m$^2$) & (GPa)
& (J/m$^2$) & (GPa)
& (J/m$^2$) & (GPa)
& (J/m$^2$) & (GPa) & \\
\midrule
GRACE-2L-OMAT-large-ft-AM   & 0.02 & 0.08 & 0.02 & 0.40 & 0.07 & 0.71 & 0.06 & 0.32 & 1.17 \\
GRACE-2L-OMAT-medium-ft-AM  & 0.04 & 0.21 & 0.03 & 0.27 & 0.10 & 0.75 & 0.10 & 0.28 & 1.39 \\
GRACE-1L-OMAT-large-ft-AM   & 0.03 & 0.29 & 0.02 & 0.22 & 0.16 & 0.97 & 0.10 & 0.48 & 1.65 \\
GRACE-1L-OMAT-medium-ft-AM  & 0.07 & 0.34 & 0.05 & 0.22 & 0.11 & 1.02 & 0.09 & 0.39 & 1.71 \\
GRACE-3L-OMAT-large-ft-AM   & 0.04 & 0.12 & 0.05 & 0.74 & 0.12 & 1.03 & 0.06 & 0.41 & 1.79 \\
GRACE-FS-OMAT              & 0.11 & 0.92 & 0.04 & 0.45 & 0.17 & 0.61 & 0.06 & 0.47 & 2.08 \\
GRACE-3L-OMAT-large        & 0.02 & 0.14 & 0.04 & 0.87 & 0.17 & 1.49 & 0.05 & 0.66 & 2.18 \\
GRACE-2L-SMAX-OMAT-medium  & 0.08 & 0.15 & 0.04 & 0.67 & 0.10 & 1.44 & 0.09 & 0.71 & 2.19 \\
GRACE-1L-SMAX-OMAT-large   & 0.03 & 0.27 & 0.04 & 0.77 & 0.15 & 1.31 & 0.08 & 0.87 & 2.32 \\
GRACE-1L-OMAT              & 0.14 & 0.14 & 0.10 & 0.41 & 0.16 & 1.14 & 0.12 & 0.33 & 2.34 \\
GRACE-2L-OMAT-large-ft-E    & 0.04 & 0.49 & 0.04 & 0.84 & 0.14 & 1.74 & 0.06 & 0.69 & 2.44 \\
GRACE-1L-OMAT-medium-ft-E   & 0.16 & 0.35 & 0.14 & 0.43 & 0.07 & 1.37 & 0.08 & 0.42 & 2.44 \\
GRACE-2L-SMAX-OMAT-large   & 0.03 & 0.37 & 0.02 & 0.81 & 0.20 & 1.64 & 0.10 & 0.92 & 2.60 \\
GRACE-1L-OMAT-large-base    & 0.03 & 0.67 & 0.02 & 1.10 & 0.16 & 1.62 & 0.10 & 0.75 & 2.72 \\
GRACE-2L-SMAX-medium       & 0.19 & 0.62 & 0.13 & 0.17 & 0.13 & 1.04 & 0.13 & 0.68 & 2.73 \\
GRACE-1L-OMAT-large-ft-E    & 0.10 & 0.45 & 0.08 & 0.86 & 0.17 & 1.69 & 0.12 & 0.51 & 2.85 \\
GRACE-2L-SMAX-large        & 0.22 & 0.92 & 0.16 & 0.48 & 0.15 & 0.64 & 0.11 & 0.54 & 2.95 \\
GRACE-1L-OAM               & 0.06 & 0.31 & 0.07 & 0.71 & 0.23 & 1.36 & 0.22 & 0.87 & 3.02 \\
GRACE-2L-OMAT-medium-ft-E   & 0.06 & 0.67 & 0.09 & 1.11 & 0.16 & 1.57 & 0.11 & 0.87 & 3.07 \\
GRACE-2L-OAM               & 0.10 & 0.57 & 0.12 & 0.77 & 0.11 & 0.89 & 0.28 & 1.25 & 3.27 \\
GRACE-1L-SMAX-large        & 0.28 & 0.77 & 0.23 & 0.39 & 0.07 & 0.86 & 0.17 & 0.79 & 3.36 \\
GRACE-2L-OMAT              & 0.12 & 0.76 & 0.15 & 1.11 & 0.20 & 1.44 & 0.11 & 0.74 & 3.44 \\
GRACE-1L-OMAT-medium-base   & 0.06 & 1.00 & 0.07 & 1.30 & 0.20 & 1.27 & 0.16 & 1.05 & 3.45 \\
GRACE-2L-OMAT-large-base    & 0.08 & 0.80 & 0.11 & 1.25 & 0.22 & 1.71 & 0.12 & 0.94 & 3.54 \\
GRACE-2L-OMAT-medium-base   & 0.15 & 0.83 & 0.18 & 1.39 & 0.21 & 1.64 & 0.17 & 1.25 & 4.32 \\
GRACE-FS-OAM               & 0.34 & 2.21 & 0.23 & 0.67 & 0.19 & 2.73 & 0.22 & 1.83 & 5.96 \\
\bottomrule
\end{tabular}
\end{table}

\clearpage

\section{Values obtained from the UBER fits to the MLIP RBS calculations}
\label{app:config_effect_all}

Table~\ref{tab:conc_stats} presents the mean values and standard deviations of $W_{\mathrm{sep}}$ and $\sigma_{\max}$ for the 100 configurations calculated per element and concentration.

\begin{table}[htbp]
\centering
\caption{Work of separation $W_{\mathrm{sep}}$ and maximum traction $\sigma_{\max}$ for each grain boundary, element and concentration (mean $\pm$ standard deviation over 100 configurations per concentration).}
\label{tab:conc_stats}
\begin{tabular}{lccccc}
\toprule
GB & Concentration (\%)
& \multicolumn{2}{c}{Cr@GB}
& \multicolumn{2}{c}{Fe@GB} \\
\cmidrule(lr){3-4}
\cmidrule(lr){5-6}
& & $W_{\mathrm{sep}}$ (J/m$^2$) & $\sigma_{\max}$ (GPa)
& $W_{\mathrm{sep}}$ (J/m$^2$) & $\sigma_{\max}$ (GPa) \\
\midrule
\multirow{5}{*}{$\Sigma 3~(111)~[\bar{1}10]$}
& 10.0 & $3.99 \pm 0.07$ & $30.02 \pm 0.39$ & $3.90 \pm 0.04$ & $29.52 \pm 0.20$ \\
& 20.0 & $4.19 \pm 0.09$ & $30.63 \pm 0.53$ & $3.95 \pm 0.04$ & $29.28 \pm 0.23$ \\
& 30.0 & $4.39 \pm 0.11$ & $31.23 \pm 0.65$ & $4.03 \pm 0.04$ & $29.18 \pm 0.21$ \\
& 40.0 & $4.55 \pm 0.13$ & $31.73 \pm 0.79$ & $4.09 \pm 0.04$ & $29.01 \pm 0.22$ \\
& 50.0 & $4.71 \pm 0.11$ & $32.29 \pm 0.66$ & $4.16 \pm 0.04$ & $28.90 \pm 0.19$ \\
\midrule
\multirow{5}{*}{$\Sigma 3~(112)~[\bar{1}10]$}
& 10.0 & $4.19 \pm 0.01$ & $28.32 \pm 0.06$ & $4.10 \pm 0.01$ & $27.89 \pm 0.05$ \\
& 20.0 & $4.38 \pm 0.02$ & $28.95 \pm 0.11$ & $4.16 \pm 0.01$ & $27.92 \pm 0.08$ \\
& 30.0 & $4.58 \pm 0.02$ & $29.67 \pm 0.15$ & $4.22 \pm 0.01$ & $27.85 \pm 0.10$ \\
& 40.0 & $4.76 \pm 0.02$ & $30.26 \pm 0.17$ & $4.27 \pm 0.01$ & $27.82 \pm 0.11$ \\
& 50.0 & $4.94 \pm 0.02$ & $30.81 \pm 0.23$ & $4.30 \pm 0.01$ & $27.68 \pm 0.13$ \\
\midrule
\multirow{5}{*}{$\Sigma 5~(012)~[100]$}
& 10.0 & $4.10 \pm 0.01$ & $26.23 \pm 0.12$ & $4.04 \pm 0.01$ & $26.08 \pm 0.05$ \\
& 20.0 & $4.22 \pm 0.02$ & $26.42 \pm 0.16$ & $4.08 \pm 0.01$ & $25.93 \pm 0.09$ \\
& 30.0 & $4.34 \pm 0.03$ & $26.60 \pm 0.18$ & $4.11 \pm 0.02$ & $25.77 \pm 0.13$ \\
& 40.0 & $4.46 \pm 0.02$ & $26.83 \pm 0.18$ & $4.14 \pm 0.02$ & $25.58 \pm 0.14$ \\
& 50.0 & $4.59 \pm 0.02$ & $27.11 \pm 0.17$ & $4.16 \pm 0.02$ & $25.38 \pm 0.13$ \\
\midrule
\multirow{5}{*}{$\Sigma 5~(013)~[100]$}
& 10.0 & $4.26 \pm 0.01$ & $27.14 \pm 0.05$ & $4.18 \pm 0.01$ & $26.81 \pm 0.07$ \\
& 20.0 & $4.39 \pm 0.01$ & $27.43 \pm 0.09$ & $4.21 \pm 0.01$ & $26.62 \pm 0.11$ \\
& 30.0 & $4.53 \pm 0.01$ & $27.75 \pm 0.10$ & $4.23 \pm 0.01$ & $26.36 \pm 0.11$ \\
& 40.0 & $4.66 \pm 0.01$ & $28.05 \pm 0.10$ & $4.23 \pm 0.02$ & $26.08 \pm 0.16$ \\
& 50.0 & $4.79 \pm 0.01$ & $28.42 \pm 0.11$ & $4.23 \pm 0.02$ & $25.72 \pm 0.15$ \\
\midrule
\multirow{5}{*}{$\Sigma 9~(221)~[\bar{1}10]$}
& 10.0 & $3.69 \pm 0.01$ & $25.31 \pm 0.07$ & $3.62 \pm 0.01$ & $24.92 \pm 0.06$ \\
& 20.0 & $3.83 \pm 0.01$ & $25.66 \pm 0.12$ & $3.65 \pm 0.01$ & $24.70 \pm 0.11$ \\
& 30.0 & $3.97 \pm 0.01$ & $26.04 \pm 0.12$ & $3.69 \pm 0.01$ & $24.48 \pm 0.12$ \\
& 40.0 & $4.10 \pm 0.01$ & $26.38 \pm 0.13$ & $3.70 \pm 0.01$ & $24.18 \pm 0.14$ \\
& 50.0 & $4.23 \pm 0.02$ & $26.69 \pm 0.13$ & $3.71 \pm 0.01$ & $23.87 \pm 0.15$ \\
\midrule
\multirow{5}{*}{$\Sigma 11~(113)~[\bar{1}10]$}
& 10.0 & $4.57 \pm 0.02$ & $29.43 \pm 0.09$ & $4.49 \pm 0.01$ & $29.07 \pm 0.06$ \\
& 20.0 & $4.76 \pm 0.02$ & $30.03 \pm 0.14$ & $4.55 \pm 0.02$ & $29.05 \pm 0.10$ \\
& 30.0 & $4.95 \pm 0.04$ & $30.65 \pm 0.20$ & $4.62 \pm 0.02$ & $29.05 \pm 0.11$ \\
& 40.0 & $5.14 \pm 0.04$ & $31.35 \pm 0.25$ & $4.67 \pm 0.02$ & $29.00 \pm 0.12$ \\
& 50.0 & $5.34 \pm 0.04$ & $32.06 \pm 0.26$ & $4.72 \pm 0.02$ & $28.96 \pm 0.12$ \\
\bottomrule
\end{tabular}
\end{table}

\end{document}